\documentclass[aip,jap,showpacs,reprint]{revtex4-1} 
\usepackage{amsmath} 
\usepackage{amsthm} 
\usepackage{amssymb}	
\usepackage{graphicx} 
\usepackage[usenames,dvipsnames]{pstricks}
\usepackage{pst-grad} 
\usepackage{pst-plot} 
\usepackage{color}
\usepackage{wasysym}
\usepackage{url}
\usepackage{bbold}
\usepackage{bm}

\usepackage[utf8]{inputenc}
\usepackage{subfigure}

\renewcommand{\v}[1]{\ensuremath{\mathbf{#1}}} 
\newcommand{\ket}[1]{\left| #1 \right>} 
\newcommand{\bra}[1]{\left< #1 \right|} 
\let\baraccent=\= 
\renewcommand{\=}[1]{\stackrel{#1}{=}} 

\theoremstyle{definition}

\theoremstyle{remark}


\begin{document}

\title{Role of Quantum Confinement in Luminescence Efficiency of Group IV Nanostructures}

\author{E. G. Barbagiovanni}
\email[]{santino.gasparo@gmail.com}
\altaffiliation{Affiliated with: Departamento de F\'{i}sica, Centro de F\'{i}sica das Interac\c{c}\~{o}es Fundamentais, Lisboa 1049-001, Portugal}
\affiliation{Laboratory for Simulation of Physical Systems, Beijing Computational Science Research Centre, Beijing 100084, People's Republic of China}

\author{D. J. Lockwood}
\affiliation{Measurement Science and Standards, National Research Council, Ottawa, Ontario K1A 0R6, Canada}

\author{N. L. Rowell}
\affiliation{Measurement Science and Standards, National Research Council, Ottawa, Ontario K1A 0R6, Canada}

\author{R. N. Costa Filho}
\affiliation{Departamento de F\'{i}sica, Universidade Federal do Cear\'{a}, Caixa Postal 6030, Campus do Pici, 60455-760 Fortaleza, Cear\'{a}, Brazil}

\author{I. Berbezier}
\affiliation{Institut Mat\'{e}riaux Micro\'{e}lectronique Nanosciences de Provence, UMR CNRS, 6137, Avenue Normandie Niemen, 13397 Marseille Cedex 20, France}

\author{G. Amiard}
\affiliation{Institut Mat\'{e}riaux Micro\'{e}lectronique Nanosciences de Provence, UMR CNRS, 6137, Avenue Normandie Niemen, 13397 Marseille Cedex 20, France}

\author{L. Favre}
\affiliation{Institut Mat\'{e}riaux Micro\'{e}lectronique Nanosciences de Provence, UMR CNRS, 6137, Avenue Normandie Niemen, 13397 Marseille Cedex 20, France}

\author{A. Ronda}
\affiliation{Institut Mat\'{e}riaux Micro\'{e}lectronique Nanosciences de Provence, UMR CNRS, 6137, Avenue Normandie Niemen, 13397 Marseille Cedex 20, France}

\author{M. Faustini}
\affiliation{Laboratoire Chimie de la Mati\`{e}re Condens\'{e}e de Paris, UMR-7574 UPMC-CNRS, Coll\`{e}ge de France, 11, place Marcelin Berthelot, 75231 Paris, France}

\author{D. Grosso}
\affiliation{Laboratoire Chimie de la Mati\`{e}re Condens\'{e}e de Paris, UMR-7574 UPMC-CNRS, Coll\`{e}ge de France, 11, place Marcelin Berthelot, 75231 Paris, France}

\date{\today}

\begin{abstract}

Experimental results obtained previously for the photoluminescence efficiency (PL$_{eff}$) of Ge quantum dots (QDs) are theoretically studied. A $\log$-$\log$ plot of PL$_{eff}$ versus QD diameter ($D$) resulted in an identical slope for each Ge QD sample only when $E_{G}\sim (D^2+D)^{-1}$. We identified that above $D\approx$ 6.2 nm: $E_{G}\sim D^{-1}$ due to a changing effective mass (EM), while below $D\approx$ 4.6 nm: $E_{G}\sim D^{-2}$ due to electron/ hole confinement. We propose that as the QD size is initially reduced, the EM is reduced, which increases the Bohr radius and interface scattering until eventually pure quantum confinement effects dominate at small $D$. 

\end{abstract}


\maketitle

\section{Introduction \label{intro}}

Early work in optoelectronic and photonic applications focused on binary semiconductor materials, because of their direct band gap \cite{Aroutiounian:2001}. Notwithstanding the indirect gap in Si and Ge, `fast' relaxation times ($\sim$ps) due to `hot' carriers \cite{deBoer:2010}, and psuedo-direct gap behaviour have been observed \cite{Barbagiovanni:2013, Averboukh:2002}. These observations along with industrial electronic fabrication compatibility have increased interest in Si and Ge applications \cite{Sueess:2013, Mangolini:2013}. Ge has a smaller gap energy ($E_G$), a larger Bohr radius \cite{Barbagiovanni:2012}, and can exhibit stronger quantum confinement (QC) effects compared to Si \cite{Cosentino:2013}. Furthermore, since Ge is compatible with Si one can realize compound structures \cite{Tsybeskov:2009}. Proper control of the electronic states with nanostructure (NS) dimension is essential for device fabrication. Theoretical studies have suggested $E_G\sim D^{-x}$, where $1\leq x\leq 2$ and $D$ is the NS diameter \cite{Niquet:2000, Barbagiovanni:2012, Barbagiovanni:2013}, which has been demonstrated experimentally \cite{Mirabella:2013, Kovalev:1998, Barbagiovanni:2013}. In this work, we theoretically elucidate the dimensional dependence (D-dep) for Ge quantum dots (QDs) using experimental results reported previously by Lockwood et al. \cite{Lockwood:2013}. We find that above $D$ $\approx$ 6 nm the QDs behave like $E_G\sim D^{-1}$, below which $E_G\sim D^{-2}$. The extra D-dep is ascribed to a change in the effective mass (EM).

\section{Experiment \label{expt}}

The Ge QDs were produced on SiO$_2$ or TiO$_2$ substrates using molecular beam epitaxy. The experimental details are in Ref. \onlinecite{Lockwood:2013}. The average QD diameter, $D$, is correlated with the original thickness of the as-deposited and subsequently annealed Ge layer. We analysed four samples: A, B, C, and D with an average Ge $D$ of 19, 21.8, 21.0, and 20.1 nm, respectively, as specified in Ref. \onlinecite{Lockwood:2013}. The size distribution with a full width half maximum of $D \approx$ 10 nm for each sample was measured using high resolution transmission electron microscopy and atomic force microscopy, and fitted with a Gaussian distribution. For sample A an amorphous Si cap was deposited in-situ atop the Ge QDs to eliminate the formation of surface related oxygen defects states. A clear observation of QC covering the weak confinement regime was noted in the experimental photoluminescence (PL) measurements for each sample.

\section{Theory \label{theory}}

\begin{figure}
\includegraphics[scale=.7]{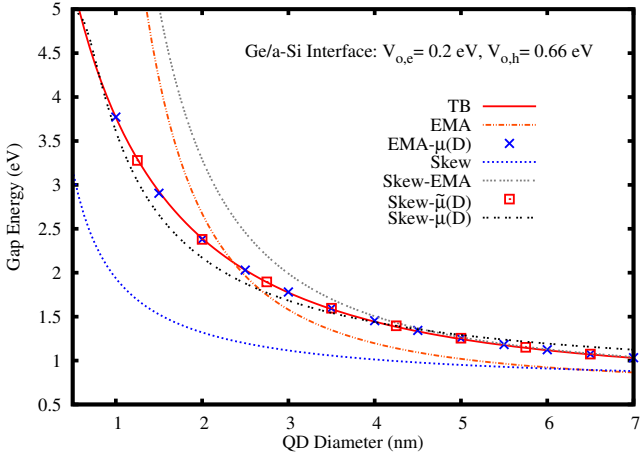}
\caption{Comparison of theoretical calculations for $E_G$ verus $D$ as described by Eqs. \eqref{TB} to \eqref{MDD}. \label{QC}}
\end{figure}

We theoretically study the PL efficiency (PL$_{eff}$), defined as the PL intensity per QD, as a function of $D$. The large near-Gaussian experimental size distribution improves the accuracy of our analysis. The PL$_{eff}$ is calculated using a range of theoretical models (Fig. \ref{QC}) and compared between the Ge QD samples. Each theoretical model is used to convert the energy axis of the PL spectrum (shown in Ref. \onlinecite{Lockwood:2013}) to $D$, then divide the PL intensity by the fitted Gaussian size distribution, which gives the PL$_{eff}$ versus $D$. First, we consider the tight binding (TB) model developed by Niquet et al. \cite{Niquet:2000}:
\begin{equation}\label{TB}
\begin{array}{ll}
&E_G(D)=E_G(\infty)+\underbrace{\frac{11.8637}{(D^{2}+2.391D+4.252)}}_{\text{CB}}+\\
&+\underbrace{\frac{15.1438}{(D^{2}+6.465D+2.546)}}_{\text{VB}}\;\text{eV};
\end{array}
\end{equation}
where $E_G(\infty)$ = 0.704 eV is the bulk gap energy corrected for optical phonon (36 meV) assisted recombination \cite{Lockwood:2013}. CB and VB denote the change in the conduction band and valence band energy, respectively. The second model is the effective mass approximation (EMA) developed by the authors \cite{Barbagiovanni:2012}: 
\begin{equation}\label{EMA}
E_G(D)=E_G(\infty)+\frac{7.88}{D^{2}}\;\text{eV}.
\end{equation}

Recently, we developed a new QC model under the assumption of a spatially dependent effective mass (SPDEM) \cite{Barbagiovanni:2013_1}, denoted throughout as the `skew' model. Briefly, we highlight a few essential features of this model. The SPDEM was derived by modifying the translation operator to include a characteristic inverse length scale, $\gamma$:
\begin{equation}\label{SPDEM}
m(x)=\frac{m_o^*}{\left(1+\gamma x\right)^2};
\end{equation}
where $x$ is the position of the charge carriers and $m_o^*$ is the bulk EM. A relation for $\gamma$ was derived:
\begin{equation}\label{gam}
\gamma^2=\frac{2}{D^2}.
\end{equation}
From Eq. \eqref{gam}, Eq. \eqref{SPDEM} expresses the EM as a function of $D$ and $x$. The net effect of the SPDEM is to reduce the carrier's EM from the bulk value (see Fig. 5 in Ref. \onlinecite{Barbagiovanni:2013_1}). We found analytic solutions to the skew model, which gives a reduced D-dep ($E_G\sim D^{-1}$) compared to the EMA:
\begin{equation}\label{skew}
E_G(D)=E_G(\infty)+
\frac{3\hbar}{\sqrt{2}D}\left[\sqrt{\frac{V_{o,e}}{m_{o,e}^*}}+\sqrt{\frac{V_{o,h}}{m_{o,h}^*}}\right];
\end{equation}
where $V_o$ is the energy difference between the QD and the matrix material at the conduction band minimum (CBM) or the valence band maximum (VBM) for an electron ($V_{o,e}$) or hole ($V_{o,h}$), respectively. $V_{o,e}$ and $V_{o,h}$ are given in Fig. \ref{QC} calculated according to the electron affinity rule for an Ge-QD surrounded by amorphous-Si (a-Si). $m_{o,e}^*$=0.56$m_{o}$ and $m_{o,h}^*$=0.29$m_{o}$ are the bulk electron and hole EM, respectively, and $m_{o}$ is the free electron mass \cite{Barbagiovanni:2012}.

\begin{figure}
\includegraphics[scale=.7]{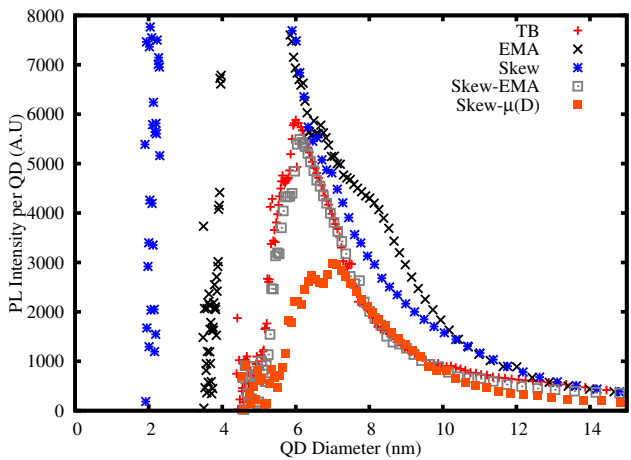}
\caption{PL$_{eff}$ versus $D$ for sample A calculated with respect to the theoretical models labeled in the figure.\label{eff}}
\end{figure}

The TB model has a D-dep of $(D^2+D)^{-1}$, while the EMA behaves like $D^{-2}$, and the skew model like $D^{-1}$. To compare these theories we consider the following modifications. First, an approximately diagonalizable Hamiltonian is given by the EMA plus the skew model (Eq. \eqref{EMA} plus \eqref{skew}), denoted `skew-EMA,' which behaves like $(D^2+D)^{-1}$. Second, in the EMA, a renormalized mass function replaces the bulk reduced EM, $\mu_o$, given by:
\begin{equation}\label{ma}
\mu(D)=\mu_o e \left(1+\frac{1}{aD^2+bD+c}\right);
\end{equation} 
where $a$, $b$, $c$, and $e$ are fitted over a range of 0.5 to 200 nm to the TB model, see Table \ref{fitpar}. We fit with the TB model, because in Ref. \onlinecite{Lockwood:2013} it resulted in a universal PL$_{eff}$ for each Ge QD sample. The fit, denoted `EMA-$\mu(D)$', is in agreement with the TB model, see Fig. \ref{QC}. Eq. \eqref{ma} also corrects the skew model, denoted `skew-$\mu(D)$', through the modification: $E_G(D)\sim (3\hbar/\sqrt{2}D\mu(D))[\cdots]$, see Table \ref{fitpar}. This fit does not show good agreement with the TB model, see Figs. \ref{QC} and \ref{eff}. However, modifying Eq. \eqref{ma} to read:
\begin{equation}\label{MDD}
\tilde{\mu}(D)=\mu_o e D\left(1+\frac{1}{aD^2+bD+c}\right);
\end{equation} 
and fitting `skew-$\tilde{\mu}(D)$' with the TB model gives agreement, see Fig. \ref{QC} and Table \ref{fitpar}. Furthermore, the fitting parameters between EMA-$\mu(D)$ and skew-$\tilde{\mu}(D)$ are in agreement, apart from $e$. The parameter $e$ in Eqs. \eqref{ma} and \eqref{MDD} is mandatory for a good fit and represents a renormalized bulk EM that is different between the TB, EMA-$\mu(D)$, and skew-$\tilde{\mu}(D)$ models.
\begin{table}
\caption{Fitting parameters for Eqs. \eqref{ma} and \eqref{MDD}. \label{fitpar}}
\begin{ruledtabular}
\begin{tabular}{c c c c} 
{} & {EMA-$\mu(D)$} & {Skew-$\tilde{\mu}(D)$} & {Skew-$\mu(D)$}\\
\hline
{$a$} & {0.047 (nm$^{-2}$)} & {0.047 (nm$^{-2}$)} & {423.463 (nm$^{-2}$)}\\
{$b$} & {0.160 (nm$^{-1}$)} & {0.160 (nm$^{-1}$)} & {-448.289 (nm$^{-1}$)}\\
{$c$} & {-0.035} & {-0.035} & {122.561}\\
{$e$} & {0.378} & {0.059 (nm$^{-1}$)} & {0.420}
\end{tabular}
\end{ruledtabular}
\end{table}

\section{Results \label{result}}

A representative calculation of the PL$_{eff}$ using sample A is shown in Fig. \ref{eff}. Skew-$\tilde{\mu}(D)$ and EMA-$\mu(D)$ results are not shown for clarity, because they lie directly atop the TB model. The peak in the PL$_{eff}$ represents the most efficient $D$ for carrier absorption and emission. Here we assume that only one electron per QD is excited in the PL experiment, which is reasonable. Whereas, the real peak in PL$_{eff}$ is associated with the volume of QDs. Here there is a sharp decline in the PL$_{eff}$ at small $D$ ($\lessapprox$ 6 nm) due to a loss of carriers \cite{Lockwood:2013}. The EMA and the skew model calculate small QD diameters down to $\approx$ 3.7 and 2 nm, respectively, which do not correlate well with experimentally measured values. The opposite behaviour is observed in the skew-$\mu(D)$ model.

\begin{figure}[ht]
 \centering
 \subfigure{
\includegraphics[scale=.7]{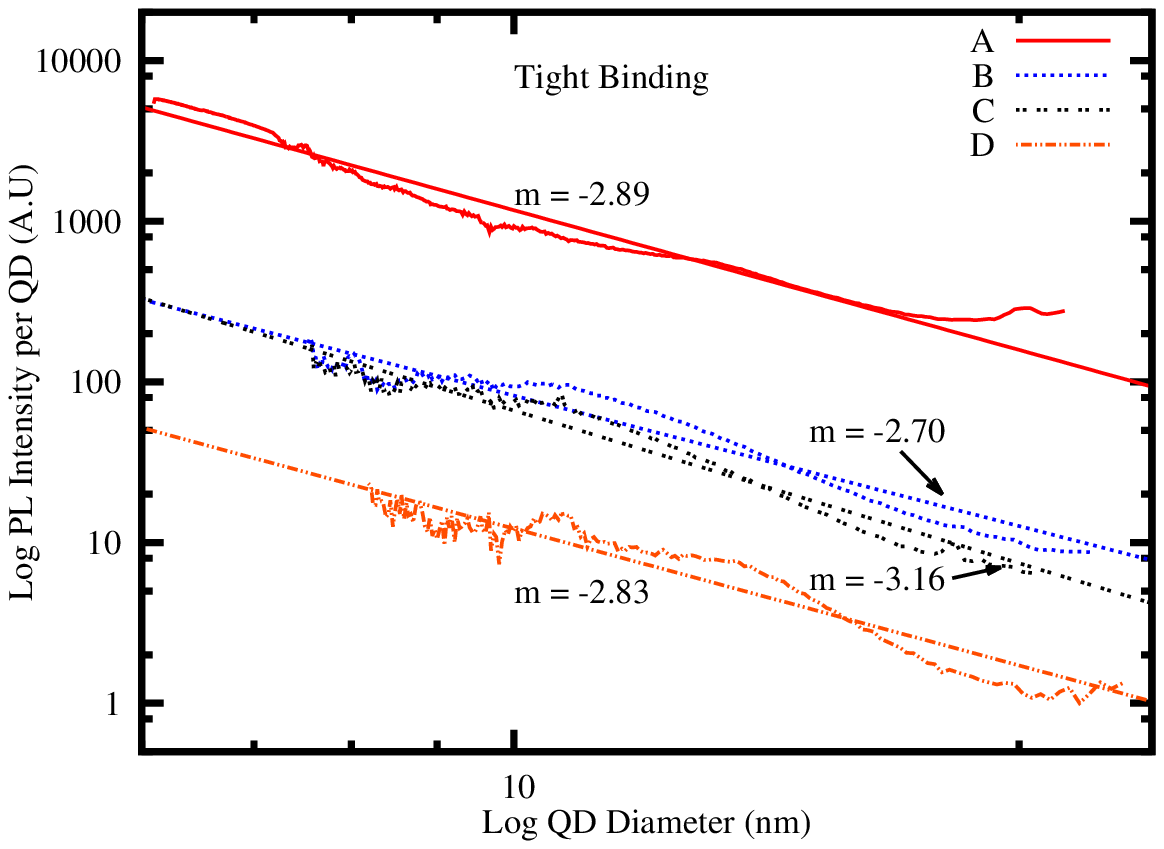}\llap{
  \parbox[b]{5in}{(a)\\\rule{0ex}{.5in}
  }}
   \label{logTB}
   }\vspace{-.3cm}
 \subfigure{
  \includegraphics[scale=.7]{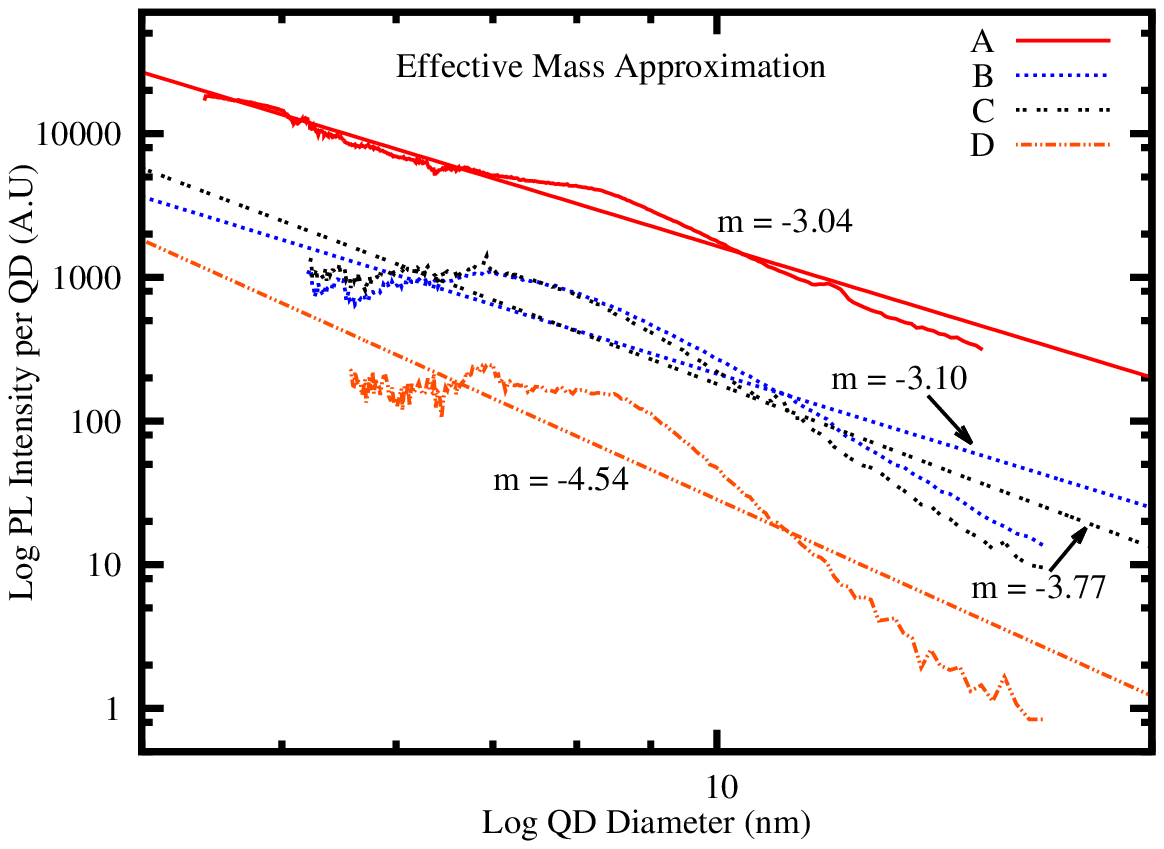}\llap{
  \parbox[b]{5in}{(b)\\\rule{0ex}{.5in}
  }}
   \label{logEMA}
   }\vspace{-.3cm}   
 \subfigure{
  \includegraphics[scale=.7]{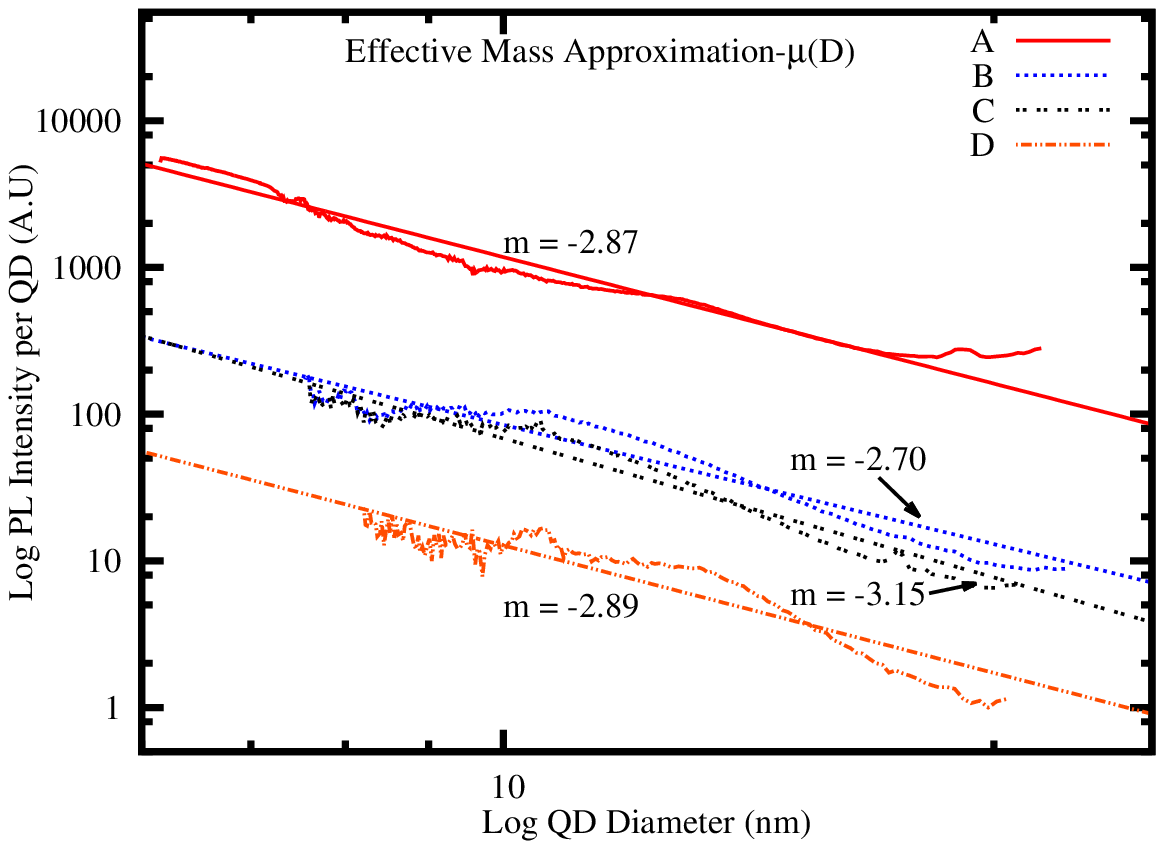}\llap{
  \parbox[b]{5in}{(c)\\\rule{0ex}{.5in}
  }}
   \label{logEMAma}
   }
\caption{Log-Log plot of PL$_{eff}$ versus $D$ calculated using \subref{logTB} Eq. \eqref{TB}, \subref{logEMA} Eq. \eqref{EMA}, and \subref{logEMAma} Eq. \eqref{ma} for the samples indicated in the figure. The linear fit ($\log(\text{PL}_{eff})=\text{a}+\text{m}\log(D)$) and m is shown for the respective samples. \label{logPLeff}}
\end{figure}

In the region $D$ $\gtrapprox$ 6 nm we observe a power law decay, which is fit to the linear equation: $\log(\text{PL}_{eff})=\text{a}+\text{m}\log(D)$, where m is the slope and a is the y-intercept, see Table. \ref{fitslope}. Note that sample B exhibits a slightly larger slope for all theories due to experimental error. Fig. \ref{logTB}, \ref{logEMA}, and \ref{logEMAma} show the degree of linear dependence in the $\log(\text{PL}_{eff})$ versus $\log(D)$ for the TB, EMA, and EMA-$\mu(D)$ models, respectively. The TB, EMA-$\mu(D)$ and skew-$\tilde{\mu}(D)$ models yield a consistent slope (m $\approx$ -2.8) for each sample, see Table. \ref{fitslope}. Similarly, the skew-EMA model (Table. \ref{fitslope}) finds nearly consistent results with the TB model, because these models share the same D-dep. On the other hand, the EMA (Fig. \ref{logEMA}) yields a large gradually increasing slope from sample A to D.
\begin{table*}
\caption{Fitting parameters and uncertantity ($\pm$m) in m from a linear fit to the experimental data: $\log(\text{PL}_{eff})=\text{a}+\text{m}\log(D)$\label{fitslope}}
\begin{ruledtabular}
\begin{tabular}{c c c c c c c c c c c c c c c c c c c c c c} 
{} & \multicolumn{3}{c}{TB} & \multicolumn{3}{c}{EMA} & \multicolumn{3}{c}{EMA-$\mu(D)$} & \multicolumn{3}{c}{Skew} & \multicolumn{3}{c}{Skew-EMA} & \multicolumn{3}{c}{Skew-$\tilde{\mu}(D)$} & \multicolumn{3}{c}{Skew-$\mu(D)$}\\ 
\hline
{} & {a} & {m} & {$\pm$m} & {a} & {m} & {$\pm$m} & {a} & {m} & {$\pm$m} & {a} & {m} & {$\pm$m} & {a} & {m} & {$\pm$m} & {a} & {m} & {$\pm$m} & {a} & {m} & {$\pm$m}\\ 
\hline
{A} & {5.96} & {-2.89} & {0.03} & {6.26} & {-3.04} & {0.02} & {5.94} & {-2.87} & {0.03} & {6.29} & {-3.11} & {0.01} & {5.99} & {-2.95} & {0.03} & {5.95} & {-2.88} & {0.02} & {6.11} & {-3.19} & {0.05}\\ 

{B} & {4.61} & {-2.70} & {0.05} & {5.43} & {-3.10} & {0.11} & {4.63} & {-2.70} & {0.06} & {5.67} & {-3.46} & {0.05} & {4.47} & {-2.64} & {0.04} & {4.63} & {-2.71} & {0.06} & {3.75} & {-2.35} & {0.08}\\ 

{C} & {4.98} & {-3.16} & {0.05} & {6.03} & {-3.77} & {0.10} & {4.98} & {-3.15} & {0.05} & {6.25} & {-4.11} & {0.05} & {4.90} & {-3.16} & {0.03} & {4.98} & {-3.15} & {0.05} & {4.21} & {-2.93} & {0.09}\\ 

{D} & {3.92} & {-2.83} & {0.07} & {3.92} & {-4.54} & {0.16} & {3.99} & {-2.89} & {0.09} & {5.46} & {-4.01} & {0.05} & {3.18} & {-2.23} & {0.05} & {3.99} & {-2.89} & {0.09} & {4.12} & {-3.33} & {0.23}\\ 
\end{tabular}
\end{ruledtabular}
\end{table*}

It can been seen from Table \ref{fitslope} that only theoretical models with a D-dep of $(D^2+D)^{-1}$ produce a universal slope (m $\approx$ -2.8) for each sample, which are the TB, EMA-$\mu(D)$, skew-$\tilde{\mu}(D)$, and skew-EMA models. To understand this behaviour we examine the essential features of each model. The EMA utilizes idealized QC conditions with an infinite confinement potential and the bulk EM values. These strong confinement conditions increase the dispersion of the confined particles where $E_G\sim D^{-2}$ dominates over $E_G\sim D^{-1}$ for small $D$. Therefore, the EMA goes to the bulk $E_G$ faster than $D^{-1}$ thus producing a small range of QD diameters, see Fig. \ref{logEMA}. The TB model was fitted to a corrected local density approximation calculation of the bulk band structure to reproduce the bulk EM and the bulk $E_G$ \cite{Niquet:2000}. These parameters are then transferred to the NS, which is terminated with H. Niquet et al. found that fitting $E_G\sim (aD^2+bD+c)^{-1}$ to their results gave much better agreement than simply using $E_G\sim D^{-2}$. Finally, the skew model utilizing an EM that varies with position and dimension produces a reduced dispersion relation: $E_G\sim D^{-1}$. The skew-EMA model can be easily corrected by considering a finite confinement potential in the EMA, which reduces the dispersion for small QD diameters \cite{Moskalenko:2007, Barbagiovanni:2013}. From this we conclude that the $D^{-1}$ dependence corresponds to a change in the EM and $D^{-2}$ corresponds to purely QC effects. 

Referring back to Table \ref{fitpar}, independent of the theoretical model, $a$ and $b$ represent a fundamental length scale of $\approx$ 4.6 nm and 6.2 nm, respectively. From these parameters, we deduce that below 4.6 nm: $E_G\sim D^{-2}$, while above 6.2 nm: $E_G\sim D^{-1}$. A tentative model is ascribed to the change in D-dep. For a large NS, when the confinement diameter is reduced and the uncertainty in momentum space increases, there is a reduction in the confined particle's EM from the bulk value,\cite{Barbagiovanni:2013_1} which scales like $D^{-1}$. The reduced EM increases the Bohr radius ($a_B\sim (m^*_o)^{-1}$) of the confined particles, thus interface scattering increases until the carriers `feel' a strong confinement potential. At which point, pure QC effects are dominant, where $E_G\sim D^{-2}$. This model is supported by recent experimental studies looking at the relationship between the interface states and pure QC effects \cite{Seguini:2013, Barbagiovanni:2012}. 

\section{Discussion \label{disc}}

Our assumption about the EM D-dep can be analysed within $\v{k}\cdot\v{p}$ perturbation theory, with\cite{Yu:2001}:
\begin{equation}\label{kpem}
\frac{1}{m_o^*}=\frac{1}{m_o}+\frac{2}{m_o^2k^2}\sum_{n'\neq n}\frac{|\bra{u_{n0}}\v{k}\cdot\v{p}\ket{u_{n'0}}|^2}{E_{n0}-E_{n'0}};
\end{equation}
where $E_{n0}$ is the energy of the $n^{th}$ band at the Brillouin zone centre, $\v{k}$ is the wave-vector, and $\v{p}$ is the momentum operator. The matrix elements ($\bra{u_{n0}}\v{k}\cdot\v{p}\ket{u_{n'0}}$) between Bloch states ($\ket{u_{n0}}$) are assumed to be given by the bulk values for a NS \cite{Cukaric:2013, Michelini:2011}. A full evaluation of Eq. \eqref{kpem} is beyond the scope of the present work. Nonetheless, we can evaluate the D-dep of Eq. \eqref{kpem}. As a trial solution, we assume that the electron and hole are confined by a Gaussian confinement potential where the envelope function in $\v{k}$-space is proportional to:
\begin{equation}\label{gauss}
F_{\v{k}}\sim\prod_{i=x,y,z}(\sigma_i^2)^{1/4} \exp{(-k_i^2\sigma_i^2/2)};
\end{equation}
where $\sigma$ is the Gaussian width and $x,y,z$ are the confinement directions. The matrix element $\bra{u_{n0}}\v{k}\cdot\v{p}\ket{u_{n'0}}$ is written in the basis of the ground state envelope function:
\begin{equation}\label{gaussem}
|\bra{n0}\v{k}\cdot\v{p}\ket{n'0}|^2\sim \sigma_i^2\int d^3\v{k}\exp{(-k_i^2\sigma_i^2/2)};
\end{equation}
where $\ket{u_{n'0}}=\ket{n'}$ and $\ket{n'0}$ denotes the ground state. From Eqs. \eqref{kpem} and \eqref{gaussem}:
\begin{equation}\label{emd}
m_o^*\sim \frac{m_o^2}{m_o+D}.
\end{equation}
Eq. \eqref{emd} is in agreement with our discussion above and with the results of our skew model. 

\section{Conclusion \label{conc}}

In conclusion, we analysed the PL obtained from a set of Ge QDs and found a universal PL$_{eff}$ if and only if the theoretical model behaves like $E_G\sim (D^2+D)^{-1}$. The universal behaviour was uncovered by comparing the D-dep between several theoretical models. The change in the D-dep of the carriers indicates that there is a change in the dominant confinement mechanism. A possible physical mechanism was ascribed to a change in the EM ($D^{-1}$), thus increasing the Bohr radius until purely QC effects dominate ($E_G\sim D^{-2}$). The samples studied here were advantageous, because they cover a large enough diameter range to observe such effects. In the case of Si NSs we have not observed this behaviour, because they are typically fabricated over a limited diameter range.  Additionally, we found that the peak of the PL intensity did not correspond well with the peak of the Gaussian size-distribution for Si QDs \cite{Iacona:2000, Saar:2005, Sias:2004}, because of defect states at the Si/SiO$_2$ interface. Nonetheless, we expect this behaviour to hold for other materials. These results help explain much of the controversy in the literature regarding the correct D-dep of $E_G$ in Si and Ge NSs and why inverse power law dependences on $D$ between 1 and 2 have been found experimentally \cite{Barbagiovanni:2013}. Furthermore, the results presented here for Ge can be utilized for device fabrication and may influence SiGe structures.

%

\end{document}